\documentclass[aps,preprint,onecolumn,11pt,showkeys]{revtex4}%
\usepackage{amsfonts}
\usepackage{amsmath}
\usepackage{amssymb}
\usepackage{graphicx}
\usepackage[ansinew]{inputenc}
\usepackage[usenames,dvipsnames]{pstricks}
\usepackage{subfigure}
\usepackage{pdfpages}
\usepackage{epsfig}
\usepackage{pst-grad}
\usepackage{pst-plot}
\usepackage[colorlinks,hyperindex]{hyperref}%
\providecommand{\U}[1]{\protect\rule{.1in}{.1in}}
\hypersetup{
colorlinks,
citecolor=blue,
linkcolor=black,
urlcolor=black,
}

\begin{document}
\title{Configurational entropy as a bounding of Gauss-Bonnet braneworld models}
\author{R. A. C. Correa$^{1}$\footnote{fis04132@gmail.com}, P. H. R. S. Moraes$^{2}%
\medskip$\footnote{moraes.phrs@gmail.com}, A. de Souza Dutra$^{1}\medskip
$\footnote{dutra@feg.unesp.br}, W. de Paula$^{2}\medskip$\footnote{wayne@ita.br},
and T. Frederico$^{2}\medskip$\footnote{tobias@ita.br}}
\affiliation{{\small {$^{1}$UNESP, Universidade Estadual Paulista, 12516-410,
Guaratinguetá, SP, Brazil}}}
\affiliation{{\small {$^{2}$ITA, Instituto Tecnológico de Aeronáutica, 12228-900, São José
dos Campos, SP, Brazil}\medskip}}
\begin{abstract}
Configurational entropy has been revealed as a reliable method for
constraining some parameters of a given model [Phys. Rev. D \textbf{92} (2015)
126005, Eur. Phys. J. C \textbf{76} (2016) 100]. In this letter we calculate
the configurational entropy in Gauss-Bonnet braneworld models. Our results
restrict the range of acceptability of the Gauss-Bonnet scalar values. In this
way, the information theoretical measure in Gauss-Bonnet scenarios opens a new
window to probe situations where the additional parameters, responsible for
the Gauss-Bonnet sector, are arbitrary. We also show that such an approach is
very important in applications that include p and Dp-branes and various
superstring-motivated theories.

\end{abstract}
\maketitle




\section{Introduction}

\label{sec:int}

In the late 90's of the last century, observational evidences of an
accelerated expansion of the universe were found
\cite{riess/1998,perlmutter/1999}. It was proposed, as an explanation for such
a counterintuitive phenomenon, the existence of the vacuum quantum energy,
which would manifest in large scales as an ``anti-gravitational force". The
vacuum quantum energy appears in the dynamical equations of the universe in
the $\Lambda$CDM (or standard) cosmological model as the cosmological constant
(CC) density parameter $\Omega_{\Lambda}$. However, when one compares the
value of $\Omega_{\Lambda}$ which can account for the present dynamical
scenario of the universe \cite{hinshaw/2013} with the value theoretically
predicted in Particle Physics \cite{weinberg/1989}, one realizes a huge
discrepancy between them, which is known as the CC problem.

An alternative to evade the CC problem - among others surrounding the standard
cosmology model, which we shall quote below - is to consider modified theories
of gravity. The $f(R)$ \cite{nojiri/2007}-\cite{nojiri/2009} and $f(R,T)$
\cite{harko/2011,ms/2016} theories of gravity have generated possibilities of
describing the cosmic acceleration with no need of a CC, evading, in this way,
the CC problem. Those theories consider the gravitational part of their action
to be dependent on a function of the Ricci scalar $R$ or of both $R$ and $T$,
with $T$ being the trace of the energy-momentum tensor.

Another possibility of describing cosmic acceleration without a CC comes from
cosmology derived from Gauss-Bonnet (GB) gravity \cite{nojiri/2005}%
-\cite{leith/2007}. Those are obtained from the consideration of the GB
invariant or a function of it in the gravitational part of the action. It is
known that higher orders in the GB term naturally arise in the low energy
limit of string theory \cite{myers/1987}.

Some alternative gravity models are a powerful tool to evade the hierarchy
problem as well, which is related to the large discrepancy among aspects of
the gravitational force and the other fundamental forces. Those are the
braneworld models \cite{antoniadis/1999}-\cite{sahni/2003}, which consider our
observable universe as a $3+1$ hypersurface (the brane) embedded in a
five-dimensional space named \textit{bulk}. Gravity, departing from the other
forces, would be able to propagate through the bulk, justifying the hierarchy
in the four-dimensional universe.

It is possible to unify some of the above formalisms in order to evade more
than one standard cosmology shortcoming simultaneously by invoking the brane
set up in $f(R)$ \cite{xu/2015}-\cite{afonso/2007}, $f(R,T)$
\cite{mc/2016,bazeia/2015b} and GB \cite{brax/2005}-\cite{feng/2009} gravity models.

The alternative gravity models mentioned above carry some ``free" parameters
with them whose values can be constrained, for instance, by cosmological
observations. Observational constraints in $f(R)$ gravity parameters can be
checked in \cite{dev/2008}-\cite{basilakos/2013}. For observational
restrictions in $f(R,T)$ and GB theories, check \cite{shabani/2014} and
\cite{he/2007}-\cite{amendola/2006}, respectively.

On the other hand, in a recent work \cite{gleiser/2012}, the concept of
entropy has been reintroduced in the literature, by taking into account the
dynamical and informational contents of models with localized energy
configurations. Based on the Shannon's information entropy, the so-called
Configurational Entropy (CE) was constructed. It can be applied to several
nonlinear scalar field models featuring solutions with spatially-localized
energy. As pointed out in \cite{gleiser/2012}, the CE can resolve situations
where the energies of the configurations are degenerate. In this case, the CE
can be used to select the best configuration. Furthermore, the authors pointed
out that this information-entropic measure is an essential tool in the study
of complex spatially-localized configurations.

We are going to discourse about the CE mechanism below. For now, it is
interesting to highlight some of its applications which reveal CE as a
powerful physical tool nowadays. For instance, it was shown in
\cite{gleiser/prd2012} that the CE quantifies the emergence of
spatially-localized, time-dependent, long-lived structures known as oscillons
\cite{osc1, osc2, osc3}. In that case, the CE is responsible for providing the
informational content of nonequilibrium field structures, in particular of
coherent states that emerge during spontaneous symmetry breaking. By
considering a Starobinsky functional form for $f(R)$, i.e., $f(R)=R+\alpha
R^{2}$, in brane models with nonconstant curvature, the $\alpha$ parameter
values were constrained by CE consideration in \cite{cmsdr/2015}. The free
parameters of the $f(R,T)=R-\alpha T$ and $f(R,T)=R+\beta R^{2}-\alpha T$
brane models were constrained in \cite{cm/2016}. It has been shown that,
indeed, CE can be used in order to extract a rich information about the
structure of the model configurations. Moreover, CE was applied to pure brane
models in \cite{correa/2015} and restrictions to the anti-de-Sitter bulk
curvature and domain wall thickness were obtained.

Studies regarding CE can also be found in solitonic $Q$-balls \cite{ent1}, in
the context of $(2+1)$-dimensional Ginzburg-Landau models \cite{ent2}, in
astrophysical objects \cite{ent3}, in two interacting scalar fields theories
\cite{correa2/2014}, in traveling solitons in Lorentz and CPT breaking systems
\cite{correa3/2015}, and in topological Abelian string-vortex and string-cigar
context \cite{correa4/2016}.

Our intention in this letter is to obtain some restrictions to the GB
braneworld parameters via CE approach. The letter is organized as follows. In
Section \ref{sec:ce} we discourse about the information content which can be
obtained from the CE approach. Since we are interested in restricting GB
braneworld models, we present a brief review of those in Section
\ref{sec:gbb}. In Section \ref{sec:ic} we calculate the CE in GB braneworld
and we discuss our results in Section \ref{sec:dc}.

\section{The configurational entropy}

\label{sec:ce}

Gleiser and Stamatopoulos (GS) \cite{gleiser/2012} have recently proposed a
detailed picture of the so-called CE for the structure of localized solutions
in classical field theories. In this section, analogously to that work, we
formulate a CE measure in the functional space, from the field configurations
where the GB braneworld scenarios can be studied.

There is an intimate link between information and dynamics, where the entropic
measure plays a prominent role. The entropic measure is well known to quantify
the informational content of physical solutions to the equations of motion and
their approximations, namely, the CE in functional space \cite{gleiser/2012}.
GS proposed that nature optimizes not solely by optimizing energy through the
plethora of \emph{a priori} available paths, but also from an informational
perspective \cite{gleiser/2012}.

The starting point is to consider structures with spatially-localized energy
and a modal fraction $f(\omega)$ which measures the relative weight of each
mode $\omega$ such that%

\begin{equation}
\label{ce1}f(\omega)=\frac{\vert\mathcal{F}\vert\omega\vert\vert^{2}}{\int
d\omega\vert\mathcal{F}\vert\omega\vert\vert^{2}},
\end{equation}
with $\mathcal{F}(\omega)$ being the Fourier transform.

The CE is defined as%

\begin{equation}
\label{ce2}\mathcal{S}_{C}(f)=-\sum f_{n}\ln f_{n}%
\end{equation}
and provides the informational content of configurations compatible with the
particular constraints of a given physical system. We can say that when all
$N$ modes carry the same weight, $f_{n}=1/N$ and the discrete CE presents a
maximum at $\mathcal{S}_{C}=\ln N$. Alternatively, if only one mode is
present, $\mathcal{S}_{C}=0$.

For general, non-periodic functions in an open interval, the continuous CE reads%

\begin{equation}
\mathcal{S}_{C}(f)=-\int d\omega\hat{f}(\omega)\ln|\hat{f}(\omega)|,
\label{ce3}%
\end{equation}
with $\hat{f}(\omega)\equiv f(\omega)/f_{max}(\omega)$ defined as the
normalized modal fraction, whereas $f_{max}(\omega)$ is the maximum fraction.
In this case, this condition ensures the positivity of $\mathcal{S}_{C}$.

\section{The Gauss-Bonnet braneworld gravity model}

\label{sec:gbb}

Since we are interesting in the calculation of the CE in GB brane models, it
is worth to briefly review such an alternative gravity theory. This can be
appreciated in the following.

An alternative to extend standard gravity is through the addition of the GB term%

\begin{equation}
\label{gbb1}G=R^{2}-4R^{\mu\nu}R_{\mu\nu}+R_{\mu\nu\lambda\rho}R^{\mu
\nu\lambda\rho},
\end{equation}
or a function of it, in the usual Einstein-Hilbert gravity lagrangian. In
(\ref{gbb1}), $R_{\mu\nu}$ is the Ricci tensor and $R_{\mu\nu\lambda\rho}$ is
the Riemann tensor. Since we are considering a five-dimensional braneworld
model, the Greek indices above assume the values $0,1,2,3,4$.

The GB brane gravity action for a general function of the GB term reads:%

\begin{equation}
S=\frac{1}{2}\int d^{4}xdy\sqrt{-g}[R+h(G)], \label{gbb2}%
\end{equation}
with $g$ being the determinant of the metric and $h(G)$ being a function of
the GB scalar.

We consider as the matter source of the universe a scalar field $\phi$
specified by the lagrangian density%

\begin{equation}
\label{gbb3}\mathcal{L}=\frac{1}{2}g_{\mu\nu}\partial^{\mu}\phi\partial^{\nu
}\phi-V(\phi)
\end{equation}
to be added to (\ref{gbb2}), with $V(\phi)$ being the scalar field potential.

Moreover we will work with the following metric%

\begin{equation}
\label{gbb4}ds^{2}=e^{2A(y)}\eta_{ab}dx^{a}dx^{b}-dy^{2},
\end{equation}
with $A(y)$ representing the warp function, $\eta_{ab}$ the Minkowski metric
and $a,b$ running from $0$ to $3$.

In \cite{bazeia/2015c}, the authors showed that for the metric (\ref{gbb4}),%

\begin{equation}
G=24(4A^{\prime\prime}A^{\prime2}+5A^{\prime4}) \label{gbb41}%
\end{equation}
and the energy-momentum tensor%

\begin{equation}
\label{gbb5}T_{ab}=\eta_{ab}\left[  \frac{1}{2}\phi^{\prime2}+V(\phi)\right]
e^{2A},
\end{equation}
with primes denoting derivatives with respect to $y$.

Now, after some straightforward calculation it is possible to obtain the
following equations of motion \cite{bazeia/2015c}:%

\begin{equation}
-3A^{\prime\prime}=2\phi^{\prime2}+12h_{G}(G)A^{\prime\prime}A^{\prime2},
\label{gbb7}%
\end{equation}%
\begin{equation}
6A^{\prime2}=\phi^{\prime2}-2V(\phi)-\frac{1}{2}h(G)+48h_{G}(G)(A^{\prime
4}+A^{\prime\prime}A^{\prime2}), \label{gbb8}%
\end{equation}
with $h_{G}(G)\equiv dh(G)/dG$. Note that within this methodology, the
potential $V(\phi)$ is a variable to be determined instead of a quantity
introduced as a basic characteristic of the theory.

For the case $h(G)=\alpha G^{n}$, with $\alpha$ and $n$ being constants, it
can be shown that the explicit form assumed by the potential is%

\begin{align}
&  \left.  V(\phi)=-\frac{3}{4}A^{\prime}+3A^{\prime2}-\frac{1}{4}\alpha
G^{n}+3\alpha nA^{\prime2}(8A^{\prime2}+7A^{\prime\prime})G^{n-1}\right.
\nonumber\\
&  \left.  -\alpha n(n-1)[A^{\prime}(14A^{\prime2}+2A^{\prime\prime}%
)G^{\prime}+(2A^{\prime2}-A^{\prime\prime})G^{\prime\prime}]G^{n-2}\right.
\nonumber\\
&  \left.  -\alpha n(n-1)(n-2)G^{\prime2}(2A^{\prime2}-A^{\prime\prime
})G^{n-3}.\right.  \label{pot}%
\end{align}

Furthermore, we have%
\begin{align}
&  \left.  \phi^{\prime2}=-\frac{3}{2}A^{\prime\prime}-6\alpha nA^{\prime
2}A^{\prime\prime}G^{n-1}\right. \nonumber\\
&  \left.  -2\alpha n(n-1)\left[  (2A^{\prime2}-A^{\prime\prime}%
)G^{\prime\prime}-2A^{\prime}(A^{\prime2}-5A^{\prime\prime})G^{\prime}\right]
G^{n-2}\right. \nonumber\\
&  \left.  -2\alpha n(n-1)(n-2)G^{\prime2}(2A^{\prime2}-A^{\prime\prime
})G^{n-3}.\right.  \label{pot1}%
\end{align}

On the other hand, from (\ref{gbb5}) the energy density $T_{00}$ is%

\begin{equation}
\rho=e^{2A}\left[  \frac{1}{2}\phi^{\prime2}+V(\phi)\right]  . \label{gbb10}%
\end{equation}

Following references \cite{bazeia/2015c,gremm/2000}, we can choose the ansatz%
\begin{equation}
A(y)=B\ln\left[  \operatorname{sech}(y)\right]  , \label{pot2}%
\end{equation}

\noindent where $B>0$.

Now, using the Eqs.(\ref{gbb41}), (\ref{pot}), (\ref{pot1}) and (\ref{pot2}),
the energy density (\ref{gbb10}) is written in the form%
\begin{equation}
\rho\left(  y\right)  =\sum\limits_{\ell=1}^{5}s_{\ell}\left(  n,B,\alpha
\right)  Q_{\ell}\left(  n,B,\alpha;y\right)  , \label{des1.1}%
\end{equation}

\noindent where we are using the following definitions%
\begin{align}
&  \left.  s_{1}\left(  n,B,\alpha\right)  \equiv\frac{3B}{2},\text{ }%
Q_{1}\left(  n,B,\alpha;y\right)  \equiv\operatorname{sech}^{2B}(y),\right. \\
& \nonumber\\
&  \left.  s_{2}\left(  n,B,\alpha\right)  \equiv-3B^{2},\text{ }Q_{2}\left(
n,B,\alpha;y\right)  \equiv\operatorname{sech}^{2B}(y)\tanh^{2}(y),\right. \\
& \nonumber\\
&  \left.  s_{3}\left(  n,B,\alpha\right)  \equiv\alpha n2^{3n}3^{n}%
B^{3n+1},\text{ }Q_{3}\left(  n,B,\alpha;y\right)  \equiv\operatorname{sech}%
^{2B}(y)\tanh^{2n+2}(y)\left[  \Psi_{B}\left(  y\right)  \right]
^{n-1},\right. \\
& \nonumber\\
&  \left.  s_{4}\left(  n,B,\alpha\right)  \equiv-\alpha n2^{3n-2}%
3^{n+1}B^{3n-1},\text{ }Q_{4}\left(  n,B,\alpha;y\right)  \equiv
\operatorname{sech}^{2B+2}(y)\tanh^{2n}(y)\left[  \Psi_{B}\left(  y\right)
\right]  ^{n-1},\right. \\
& \nonumber\\
&  \left.  s_{5}\left(  n,B,\alpha\right)  \equiv-\alpha2^{3n-2}3^{n}%
B^{3n},\text{ }Q_{5}\left(  n,B,\alpha;y\right)  \equiv\operatorname{sech}%
^{2B}(y)\tanh^{2n}(y)\left[  \Psi_{B}\left(  y\right)  \right]  ^{n},\right.
\end{align}

\noindent with%
\begin{equation}
\Psi_{B}\left(  y\right)  \equiv5B\tanh^{2}(y)-4\operatorname{sech}^{2}(y).
\end{equation}

The profiles for the energy density and warp factor are depicted in Fig.1,
which shows the influence of the $B$ parameter on the configurations. It is
important to explain the reason for working with small values of B. In Fig.1,
the reader can note that when $B$ increases the brane is narrowed down.
Moreover, simultaneously, the energy density develops lateral peaks and a
central valley which goes rapidly to zero as $B$ increases. Such a critical
phenomenon of thick brane models is called \textquotedblleft brane splitting"
and it was first presented in \cite{campos/2002}. It has already appeared in
GB \cite{bazeia/2015} and $f(R)$ \cite{bazeia/2014} models. Since in the brane
splitting case, the field will not be confined to the brane, we avoid large
values of $B$.


\begin{figure}[h]
\includegraphics[scale=1.3]{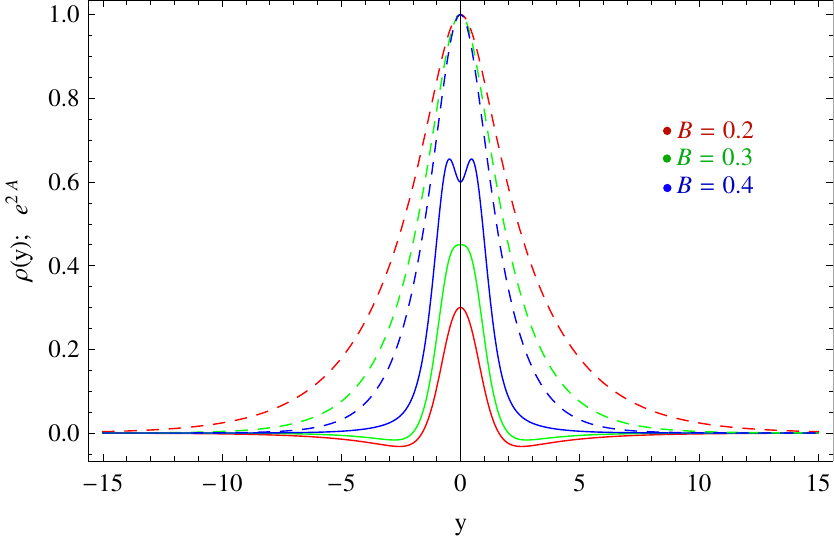}\caption{Energy density (thin continuous
line) and warp factor (dashed line) with $\alpha=1$ and $n=1$.}%
\end{figure}


\section{Information content in Gauss-Bonnet braneworld}

\label{sec:ic}

Now we will apply the new concept of CE in the functional space from the field
configurations where the GB braneworld scenarios can be studied. To begin, we
write the Fourier transform%
\begin{equation}
\mathcal{F}[\omega]=\frac{1}{\sqrt{2\pi}}\int dy\;e^{i\omega y}\rho(y),
\label{3.1}%
\end{equation}
where $\rho(y)$ is the standard energy density. Here, it is important to
remark that the energy density given by Eq.(\ref{des1.1}) is localized. In
fact, as shown by GS, the condition necessary so that the CE is well-defined
in physical applications is that the energy density is represented by a
spatially-confined structure.

By using the Plancherel theorem, it follows that%
\begin{equation}
\int d\omega\left\vert \mathcal{F}[\omega]\right\vert ^{2}=\int dr\left\vert
\rho(y)\right\vert ^{2}. \label{3.1.1}%
\end{equation}

Now, substituting the energy density given by Eq.(\ref{des1.1}) into
Eq.(\ref{3.1}), we can obtain, after some arduous calculation, the following
Fourier transform
\begin{equation}
\mathcal{F}(\omega)=\sum\limits_{\ell=1}^{5}\sum\limits_{j=1}^{2}\tilde
{s}_{\ell}\left(  n,B,\alpha\right)  \mathcal{I}_{\ell,j}\left(
n,B,\alpha;\omega\right)  , \label{eq.4}%
\end{equation}
where $\tilde{s}_{\ell}\left(  n,B,\alpha\right)  \equiv s_{\ell}\left(
n,B,\alpha\right)  /\sqrt{2\pi}$ and $\mathcal{I}_{\ell,j}[n,B,\alpha;\omega]$
are the following corresponding functions%

\begin{align}
&  \left.  \mathcal{I}_{1,1}\left(  n,B,\alpha;\omega\right)  \mathcal{=}%
\frac{4}{c+i\omega}\mathcal{G}\left[  c,\frac{c+i\omega}{2};\frac{c+i\omega
}{2}+1;-1\right]  ,\right. \\
& \nonumber\\
&  \left.  \mathcal{I}_{1,2}\left(  n,B,\alpha;\omega\right)  \mathcal{=}%
\frac{4}{c-i\omega}\mathcal{G}\left[  c,\frac{c-i\omega}{2};\frac{c-i\omega
}{2}+1;-1\right]  ,\right.
\end{align}

\begin{align}
&  \left.  \mathcal{I}_{2,1}\left(  n,B,\alpha;\omega\right)  \mathcal{=}%
\frac{1}{2+\tilde{b}_{0}+i\omega}\mathcal{G}\left[  \tilde{b}_{0}%
,\frac{2+\tilde{b}_{0}+i\omega}{2};\frac{4+\tilde{b}_{0}+i\omega}%
{2};-1\right]  \right. \nonumber\\
&  \left.  +\frac{1}{b_{0}-i\omega}\mathcal{G}\left[  \tilde{b}_{0}%
,\frac{b_{0}-i\omega}{2};\frac{b_{0}-i\omega}{2}+1;-1\right]  ,\right.
\end{align}
\begin{align}
&  \left.  \mathcal{I}_{2,2}\left(  n,B,\alpha;\omega\right)  \mathcal{=}%
\frac{1}{b_{0}+i\omega}\mathcal{G}\left[  \tilde{b}_{0},\frac{b_{0}+i\omega
}{2};\frac{b_{0}+i\omega}{2}+1;-1\right]  \right. \nonumber\\
&  \left.  +\frac{1}{i\omega-2-\tilde{b}_{0}}\mathcal{G}\left[  \tilde{b}%
_{0},\frac{i\omega-2-\tilde{b}_{0}}{2};\frac{i\omega-\tilde{b}_{0}}%
{2};-1\right]  ,\right. \\
& \nonumber\\
&  \left.  \mathcal{I}_{3,1}\left(  n,B,\alpha;\omega\right)  \mathcal{=}%
\frac{\left(  \alpha_{1}\alpha_{2}\right)  ^{n-1}\Gamma\lbrack\gamma
_{1}+1]\Gamma\lbrack\gamma_{2}+1]}{2\Gamma\lbrack\gamma_{1}+\gamma_{2}%
+2]}\mathcal{H}_{D}^{(3)}\left[  \gamma_{1}+1;\gamma_{3},-\gamma_{4}%
,-\gamma_{4};\gamma_{1}+\gamma_{2}+2;-1,v,u\right]  ,\right. \\
& \nonumber\\
&  \left.  \mathcal{I}_{3,2}\left(  n,B,\alpha;\omega\right)  \mathcal{=}%
\frac{\left(  \alpha_{1}\alpha_{2}\right)  ^{n-1}\Gamma\lbrack1-\bar{\gamma
}_{0}]\Gamma\lbrack\bar{\gamma}_{1}+1]}{2\Gamma\lbrack\bar{\gamma}_{1}%
-\bar{\gamma}_{0}+2]}\mathcal{H}_{D}^{(3)}\left[  1-\bar{\gamma}_{0}%
;\bar{\gamma}_{2},-\bar{\gamma}_{3},-\bar{\gamma}_{4};\bar{\gamma}_{1}%
-\bar{\gamma}_{0}+2;-1,v,u\right]  ,\right. \\
& \nonumber\\
&  \left.  \mathcal{I}_{4,1}\left(  n,B,\alpha;\omega\right)  =\frac
{2^{\delta-1}\xi_{1}^{\theta}\left(  \epsilon_{1}\epsilon_{2}\right)
^{\theta}\Gamma\lbrack R]\Gamma\lbrack2\zeta+R+1]}{\Gamma\lbrack2\zeta
+R+1]}\mathcal{H}_{D}^{(3)}\left[  R;\Phi,-\theta,-\theta;2\zeta
+R+1;-1,\tilde{v},\tilde{u}\right]  ,\right. \\
& \nonumber\\
&  \left.  \mathcal{I}_{4,2}\left(  n,B,\alpha;\omega\right)  =\frac
{2^{\tilde{\delta}-1}\tilde{\xi}_{1}^{\tilde{\theta}}\left(  \epsilon
_{1}\epsilon_{2}\right)  ^{\tilde{\theta}}\Gamma\lbrack\tilde{R}]\Gamma
\lbrack\tilde{\lambda}+1]}{\Gamma\lbrack\lambda+\tilde{R}+1]}\mathcal{H}%
_{D}^{(3)}\left[  \tilde{R};\Phi,-\theta,-\theta;\lambda+R+1;-1,\hat{v}%
,\hat{u}\right]  ,\right. \\
& \nonumber\\
&  \left.  \mathcal{I}_{5,1}\left(  n,B,\alpha;\omega\right)  =\frac
{2^{\hat{\delta}-1}\hat{\xi}_{1}^{\theta}\left(  \tilde{\epsilon}_{1}%
\tilde{\epsilon}_{2}\right)  ^{\theta}\Gamma\lbrack\hat{R}]\Gamma\lbrack
2\hat{\zeta}+\hat{R}+1]}{\Gamma\lbrack2\zeta+R+1]}\mathcal{H}_{D}^{(3)}\left[
\hat{R};\hat{\Phi},-\hat{\theta},-\hat{\theta};2\hat{\zeta}+\hat
{R}+1;-1,U,Y\right]  ,\right. \\
& \nonumber\\
&  \left.  \mathcal{I}_{5,2}\left(  n,B,\alpha;\omega\right)  =\frac
{2^{\mathring{\delta}-1}\mathring{\xi}_{1}^{\mathring{\theta}}\left(
\tilde{\epsilon}_{1}\tilde{\epsilon}_{2}\right)  ^{\tilde{\theta}}%
\Gamma\lbrack\mathring{R}]\Gamma\lbrack\mathring{\lambda}+1]}{\Gamma
\lbrack\lambda+\mathring{R}+1]}\mathcal{H}_{D}^{(3)}\left[  \mathring
{R};\mathring{\Phi},-\mathring{\theta},-\mathring{\theta};\lambda+\mathring
{R}+1;-1,U,Y\right]  ,\right.
\end{align}

\noindent with the definitions%
\begin{align*}
&  \left.  c\equiv2B+2,\text{ }b_{0}\equiv2B,\text{ }\tilde{b}_{0}%
\equiv2B+2,\text{ }\alpha_{1}\equiv\frac{d_{0}}{2}+\sqrt{\frac{d_{0}^{2}}%
{4}-1},\text{ }\alpha_{2}\equiv\frac{d_{0}}{2}-\sqrt{\frac{d_{0}^{2}}{4}%
-1},\right. \\
&  \left.  d_{0}\equiv\frac{5B}{2}+4,\text{ }\epsilon_{1}\equiv\frac{D_{0}}%
{2}+\sqrt{\frac{D_{0}^{2}}{4}-1},\text{ }\epsilon_{2}\equiv\frac{D_{0}}%
{2}-\sqrt{\frac{D_{0}^{2}}{4}-1},\text{ }D_{0}\equiv2+\frac{4}{5B},\right. \\
&  \left.  \gamma_{1}\equiv\frac{2B-4n+4+i\omega}{2}-1,\text{ }\gamma
_{2}\equiv2n+2,\text{ }\gamma_{3}\equiv2B+4,\text{ }\gamma_{4}\equiv
n-1,\text{ }v\equiv\frac{1}{\alpha_{1}},\text{ }u\equiv\frac{1}{\alpha_{2}%
},\right. \\
&  \left.  \bar{\gamma}_{0}\equiv\frac{2B-4n+4-i\omega}{2}-1,\text{ }%
\bar{\gamma}_{1}\equiv2n+2,\text{ }\bar{\gamma}_{2}\equiv-2B-4,\text{ }%
\bar{\gamma}_{3}=\bar{\gamma}_{4}\equiv n-1,\text{ }\xi_{1}\equiv5B,\right. \\
&  \left.  \zeta\equiv n,\text{ }\theta\equiv n-1,\text{ }\Phi\equiv
2\theta+2B+2+2n,\text{ }R\equiv i\omega-2\zeta-2\theta+\Phi,\text{ }%
U\equiv\frac{1}{\tilde{\epsilon}_{2}},\text{ }Y\equiv\frac{1}{\tilde{\epsilon
}_{1}},\right. \\
&  \left.  \delta\equiv2B+2,\text{ }\tilde{v}=\hat{v}\equiv\frac{1}%
{\epsilon_{1}},\text{ }\tilde{u}=\hat{u}\equiv\frac{1}{\epsilon_{2}},\text{
}\tilde{R}=-\hat{R}\equiv i\omega-2\zeta+2\theta+\Phi,\text{ },\right. \\
&  \left.  \hat{\Phi}=-\mathring{\Phi}\equiv2\theta-2B+2+2n,\text{ }%
\hat{\theta}=-\mathring{\theta}\equiv2n-1,\text{ }\mathring{R}\equiv
-i\omega-2\zeta+2\theta+\Phi.\right.
\end{align*}

Furthermore, in the above expressions, $\mathcal{G}\left[  \odot,\odot
;\odot;\odot\right]  $ stand for the well-known hypergeometric functions and
$\mathcal{H}_{D}^{(3)}\left[  \otimes;\otimes,\otimes;\otimes;\otimes
,\otimes,\otimes\right]  $ is the so-called Lauricella functions of three
variables \cite{lauricella}.

Thus, the modal fraction (\ref{ce1}) becomes
\begin{equation}
f(\omega)=\frac{\sum_{\ell,\ell^{\prime}=1}^{5}\sum_{j,j^{\prime}=1}^{2}%
\tilde{s}_{\ell}\left(  n,B,\alpha\right)  \tilde{s}_{\ell^{\prime}}^{\ast
}\left(  n,B,\alpha\right)  \mathcal{I}_{\ell,j}\left(  n,B,\alpha
;\omega\right)  \mathcal{I}_{\ell^{\prime},j^{\prime}}^{\ast}\left(
n,B,\alpha;\omega\right)  }{\sum_{\ell,\ell^{\prime}=1}^{5}\sum_{j,j=1}%
^{2}\tilde{s}_{\ell}\left(  n,B,\alpha\right)  \tilde{s}_{\ell^{\prime}}%
^{\ast}\left(  n,B,\alpha\right)  \int d\omega\mathcal{I}_{\ell,j}\left(
n,B,\alpha;\omega\right)  \mathcal{I}_{\ell^{\prime},j^{\prime}}^{\ast}\left(
n,B,\alpha;\omega\right)  }. \label{3.9}%
\end{equation}

In Fig.2 the modal fraction is depicted for different values of the parameter
$B$.

\begin{figure}[h]
\center
\subfigure[ref1][\hspace{0.3cm}Modal fraction with $n=1$ and $\alpha=1$.]{\includegraphics[width=7.8cm]{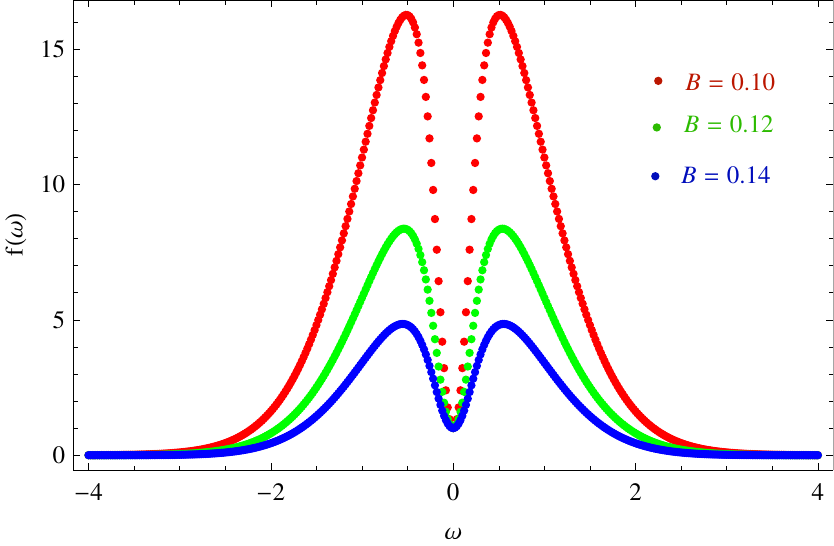}}
\qquad
\subfigure[ref2][\hspace{0.3cm}Modal fraction with $n=2$ and $\alpha=1$.]{\includegraphics[width=7.8cm]{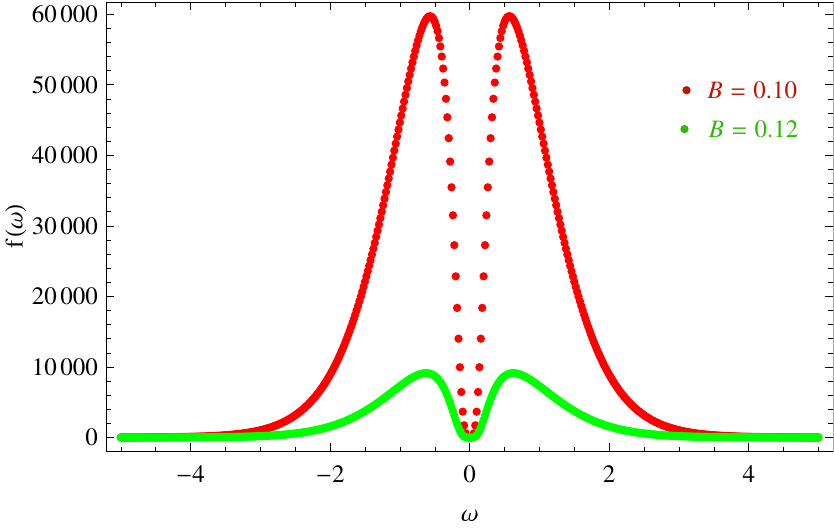}}\caption{Modal
fractions}%
\end{figure}

In Fig.3 it can be seen that given a value of $B$ there is a value of $\alpha$
for which CE is a minimum. Furthermore when $B$ decreases, that minimum value
also decreases. This results are in agreement with the CE concept found in
\cite{gleiser/2012}. Here, we can observe that lower CE correlates with lower
energy. Thus, the most prominent solutions to the GB theory under analysis are
given for some specific values of $\alpha$. This will be analysed further in
the next section.

\begin{figure}[h]
\center
\subfigure[ref1][\hspace{0.3cm}CE with $n=1$.]{\includegraphics[width=7.8cm]{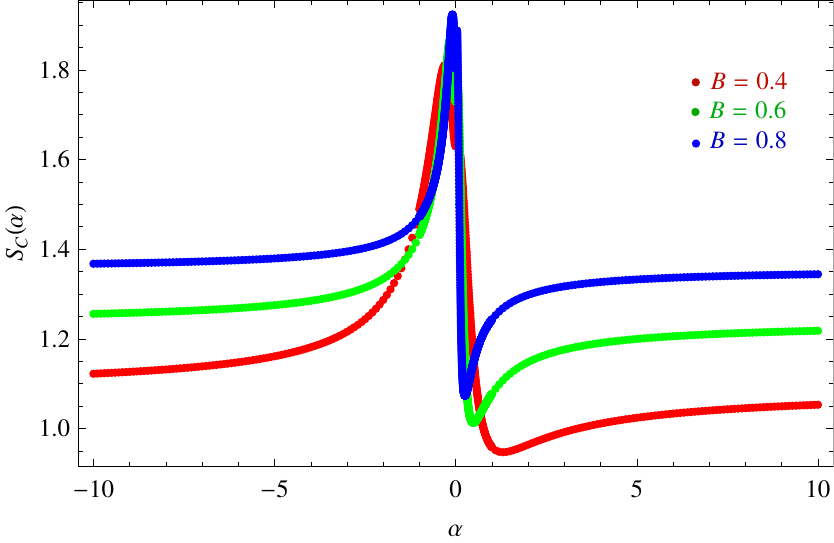}}
\qquad
\subfigure[ref2][\hspace{0.3cm}CE with $n=2$.]{\includegraphics[width=7.8cm]{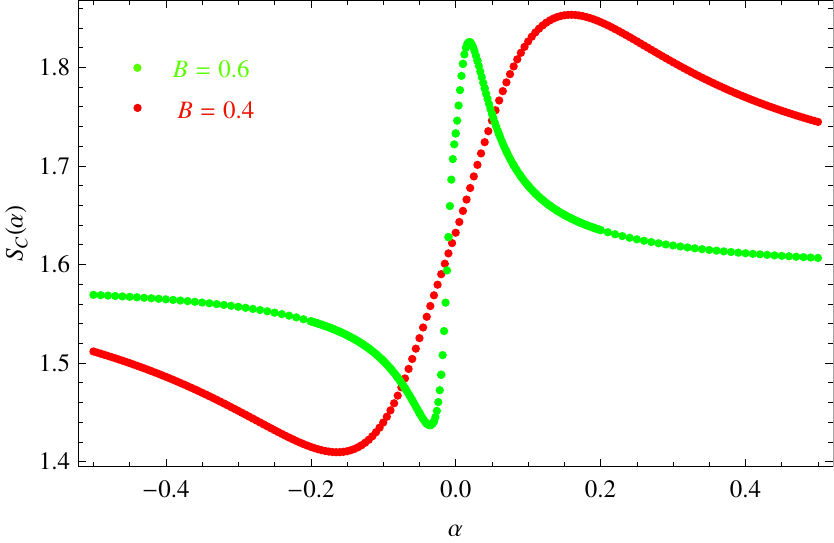}}\caption{Configurational
entropy}%
\end{figure}

\section{Discussion and conclusions}

\label{sec:dc}

The CE has been revealed as a reliable method for constraining some given
model parameters, as one can check, for instance, in \cite{cmsdr/2015,cm/2016}%
. The entropic information has been here studied in braneworld models, with
emphasis on the GB scenario, which has been chosen by its very physical
content and usefulness. We have shown that the information theoretical measure
of GB braneworld models opens new possibilities to physically constrain, for
example, parameters that are related to the GB term. The CE provides the most
appropriate value of these parameters that are consistent with the best
organizational structure.

The information measure of the system organization is related to modes in the
braneworld model. Hence the constraints of the parameters that we obtained for
the GB model provide the range of the parameters associated to the most
organized braneworld models with respect to the information content of these models.

By analysing Fig.3 of the previous section, we can see that for high values of
$B$, the CE minimum, which is related to the reliable solutions of the system,
is found for $\alpha=0$ or $\alpha\simeq0$. In other words, there is a specific value for $B$ which makes the GB term to vanish. Therefore, the CE serves as a new tool to specify the dynamic constraints of the GB model.

As we can see, the CE provides a complementary perspective to investigate
alternative theories of gravity. Further interests that concerns to CE and
which we are interested in, can be found in dynamical bound in alternatives
theory of gravity, such as Brans-Dicke \cite{brans/1961}, Kaluza-Klein
\cite{overduin/1997} and $f(R,L_{m})$ \cite{harko/2010} gravity models, among
many others.

\bigskip


\bigskip

\begin{acknowledgments}
RACC thanks CAPES for financial support and to the Physics Graduate Program at
Unesp-Campus Guaratinguetá for hospitality. PHRSM would like to thank S\~ao
Paulo Research Foundation (FAPESP), grant 2015/08476-0, for financial support.
ASD and WP are thankful to the CNPq for financial support. TF is thankful to the CNPq, FAPESP, and CAPES.
\end{acknowledgments}

\end{document}